\documentclass[prb,aps,twocolumn,superscriptaddress,floatfix,showpacs,longbibliography,reprint]{revtex4-2}
\usepackage{graphicx}
\usepackage{hyperref}
\usepackage{dcolumn}
\usepackage{bm}
\usepackage{upgreek}
\usepackage[normalem]{ulem}
\usepackage{amsmath,amssymb}
\usepackage{dsfont}
\usepackage[english]{babel}
\usepackage{braket}
\usepackage{diagbox}
\usepackage{color}

\begin{document}
\title{Interaction-induced moir\'e lattices: from mosaic mobility edges to many-body localization}
\author{Yan-Hao Yang}
\affiliation{Guangdong Basic Research Center of Excellence for Structure and Fundamental Interactions of Matter, Guangdong Provincial Key Laboratory of Quantum Engineering and Quantum Materials, School of Physics, South China Normal University, Guangzhou 510006, China}
\author{Zhihao Xu}
\affiliation{Institute of Theoretical Physics and State Key Laboratory of Quantum Optics Technologies and Devices, Shanxi University, Taiyuan, 030006, China}
\author{Lei Ying}
\affiliation{School of Physics and Zhejiang Key Laboratory of Micro-nano Quantum Chips and Quantum Control, Zhejiang University, Hangzhou 310027, China}
\author{Qizhong Zhu}
\email{qzzhu@m.scnu.edu.cn}
\affiliation{Guangdong Basic Research Center of Excellence for Structure and Fundamental Interactions of Matter, Guangdong Provincial Key Laboratory of Quantum Engineering and Quantum Materials, School of Physics, South China Normal University, Guangzhou 510006, China}
\affiliation{Guangdong-Hong Kong Joint Laboratory of Quantum Matter, Frontier
Research Institute for Physics, South China Normal University, Guangzhou
510006, China}

\date{\today}

\begin{abstract}
We study localization driven solely by interparticle interactions in moir\'e lattice systems without intrinsic disorder or externally imposed quasiperiodic potentials. We consider a one-dimensional bilayer with incommensurate lattice constants, described by a spin-dependent Fermi-Hubbard-type model with short-range interlayer interactions, where quasiperiodicity emerges only through interactions. Exact diagonalization shows that quenching hopping in one layer generates an interaction-induced mosaic potential with multiple mobility edges. When both layers are dynamical, increasing interlayer interactions drives transitions among ergodic, critical, and many-body localized regimes, with energy-dependent coexistence in certain parameter ranges. An exact mapping to a noninteracting single-particle model on a higher-dimensional structured graph provides a unified interpretation of these results and suggests an experimentally accessible route to interaction-induced moir\'e physics and localization. 
\end{abstract}

\maketitle

\section{Introduction}

Moir\'e superlattices are increasingly viewed as a promising method for probing quantum matter in condensed‑matter systems \cite{kennes2021moire}. A small twist angle or lattice mismatch between two periodic layers generates long-wavelength patterns that reconstruct the electronic spectrum and can produce narrow or nearly flat bands, thereby amplifying correlation effects \cite{Andrei2021Marvels,kennes2021moire}. This setting has enabled the discovery of a broad range of emergent phases, spanning correlated insulators and unconventional superconductivity in magic-angle graphene \cite{Cao2018CorrelatedInsulator,Cao2018UnconventionalSC}, as well as topology-enabled correlated phases such as moir\'e Chern insulators and the quantized anomalous Hall (QAH) effect \cite{Sharpe2019QAH,Serlin2020QAH}. More recently, moir\'e semiconductors have pushed this direction further by realizing fractional topological states in flat moir\'e bands \cite{Park2023FQAH,Kang2024FQSH,zeng2023thermodynamic,xu2023observation,cai2023signatures,lu2024fractional}. Motivated by these advances, substantial efforts have also been devoted to simulating and controlling moir\'e physics with ultracold atoms, where geometry, twist/mismatch, interactions, and filling can be tuned flexibly. In particular, twisted-bilayer optical lattices have now been demonstrated experimentally \cite{Meng2023TwistedBilayerBEC,yao2020many,sierant2018many,sbroscia2020observing,sierant2022challenges,guoObservationEnergyresolvedManybody2021} and used to access moir\'e minibands, localization and correlation-driven phenomena, providing a clean and highly controllable platform complementary to solid-state moir\'e materials \cite{Meng2023TwistedBilayerBEC,Wang2024ThreeDMoireCrystal,gonzalez2019cold,salamon2020simulating,luo2021spin,paul2023particle,madronero2023dynamic,zeng2025interaction,ding2025interaction,li2024ground,zhang2025dipolar,bloch2008many,chin2010feshbach,gross2021quantum,PhysRevA.111.023320,PhysRevB.100.144202}.

A complementary line of research in moir\'e and, more broadly, incommensurate systems concerns localization induced by quasiperiodicity. Deterministic incommensurate potentials can localize single-particle eigenstates even in the absence of true randomness, as exemplified by the Aubry--Andr\'e mechanism and its extensions. Cold-atom experiments have played a prominent role here, offering direct observations of localization in quasiperiodic lattices and enabling systematic probes of mobility-edge physics in generalized settings \cite{Roati2008AndersonQP}. More recently, quasicrystalline optical lattices have extended this program to higher-dimensional quasiperiodic potentials, where localization can be studied without disorder \cite{Viebahn2019Quasicrystal,Sbroscia2020QuasicrystalLocalization,yu2024observing}. Importantly, these developments primarily address single-particle localization (or weakly interacting limits). Once interactions are appreciable, the key question becomes whether localization can persist at finite energy density, i.e., whether many-body localization (MBL) can occur. Indeed, MBL and slow relaxation have been extensively investigated in quasiperiodic lattices, including in cold-atom realizations \cite{Luschen2017QPMBL,Bordia2017QPMBL2D}. In essentially all of these studies, however, quasiperiodicity is an explicit single-particle ingredient, and interactions act on top of an externally imposed aperiodic landscape.

This observation motivates a distinct route to quasiperiodicity, in which the effective aperiodic structure is generated by interactions themselves. In interaction-induced moir\'e lattices, two otherwise periodic layers are coupled solely through interlayer interactions, so that the spatial modulation of interaction energy acts as an emergent moir\'e potential without introducing external disorder or single-particle quasiperiodicity \cite{zeng2025interaction}. Existing work on this interaction-generated moir\'e mechanism has largely focused on mean-field and ground-state physics, predicting unconventional superfluid and Mott regimes, symmetry-breaking states tied to the emergent moir\'e pattern, and Bose-glass--like behavior without extrinsic disorder \cite{zeng2025interaction}. This leaves open a deeper, genuinely many‑body question at the heart of localization physics: can interparticle interactions—beyond any mean‑field treatment—together with incommensurability, generate robust localization, including in highly excited eigenstates, even when the underlying single‑particle system is perfectly periodic?

In this work we address this gap by exploring the full many-body spectrum of interaction-induced moir\'e systems, with an emphasis on localization phenomena. The key question we ask is whether interparticle interactions, combined with lattice incommensurability, can induce localization across the spectrum and under what conditions such localization survives at finite energy density. Our setting differs qualitatively from conventional quasiperiodic MBL models: in the absence of interlayer interactions, each layer is strictly periodic and hence all states should be extended. Any effective quasiperiodicity arises only through interlayer coupling and is therefore intrinsically interaction generated.

To make the problem tractable while retaining the essential physics, we study a spin-dependent bilayer lattice described by a Fermi-Hubbard-type model with incommensurate lattice constants for the two components. Interlayer tunneling is neglected, while short-range interlayer density-density interactions couple particles residing in different lattices. This minimal setup captures the core ingredient of interaction-induced moir\'e physics and is well suited for exact diagonalization. Importantly, it is also experimentally realistic in cold-atom platforms \cite{schreiber2015observation,bordia2016coupling,islam2015measuring,altman2015universal,nandkishore2015many}, including optical lattices with species-dependent wavelengths and optical tweezer arrays \cite{PhysRevLett.128.033201,PhysRevX.8.041055,PhysRevX.8.041054,PhysRevLett.122.143002}, where hopping and interaction scales can be tuned over wide ranges.

Within this framework, we identify two complementary routes to localization. First, by freezing the dynamics in one layer, particles in the other layer experience an interaction-induced quasiperiodic potential with a mosaic-like envelope, leading to single-particle localization with multiple mobility edges. Second, when hopping is allowed in both layers, the same interaction-generated quasiperiodicity acts in the full many-body Hilbert space and can drive transitions among ergodic, critical, and many-body localized regimes. The resulting phases depend strongly on filling and interaction strength: in particular, strong interlayer interactions can induce localization at high fillings, while low-filling regimes remain largely extended or thermal.

Together, these results establish interaction-induced moir\'e lattices as a disorder-free and experimentally accessible platform for studying localization physics. They also provide a concrete realization of the counterintuitive notion that interactions alone can serve as the driving mechanism for localization. In the next section, we introduce the microscopic lattice model and discuss its experimental tunability in detail.

The remainder of this paper is organized as follows. In Sec.~II, we introduce the spin-dependent bilayer lattice model with incommensurate lattice constants and short-range interlayer interactions, and discuss its experimental realizability and parameter tunability. In Sec.~III, we consider the quenched-layer limit, where hopping in one layer is suppressed; we show that the remaining layer experiences an interaction-induced mosaic quasiperiodic potential and exhibits single-particle localization with multiple mobility edges. Section~IV addresses the fully dynamical case with hopping in both layers: using participation entropies and level statistics, we demonstrate interaction-induced MBL and map out the phase diagram as a function of filling and interaction strength. In Sec.~V, we present an exact mapping of the interacting many-body problem to a noninteracting single-particle model on a higher-dimensional structured graph, providing a unified interpretation of the observed localization phenomena. Finally, Sec.~VI summarizes our results and discusses extensions to higher-dimensional and twisted moir\'e lattice systems, as well as experimental prospects.

\begin{figure}[h]
   \centering
           \includegraphics[width=1\hsize,keepaspectratio]{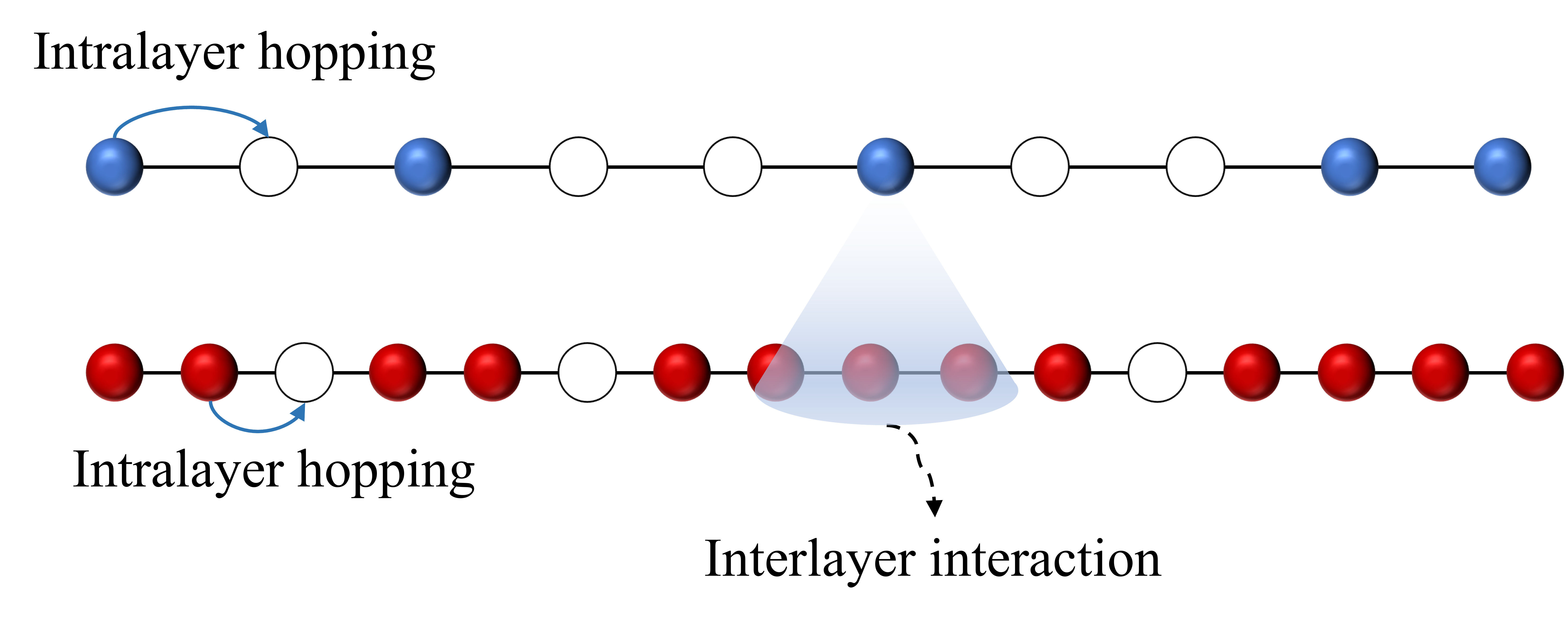}
       \caption{Schematic of the bilayer lattice model for $N_{\uparrow}=4$ on a lattice of size $L_{\uparrow}=10$ and $N_{\downarrow}=13$ on a lattice of size $L_{\downarrow}=16$. Particles hop between nearest-neighbor sites within each layer, while the two layers are coupled through short-range interlayer density-density interactions.}
    \label{fig: figure1}
\end{figure}

\section{Model Hamiltonian and experimental realization}

We consider a spin-dependent bilayer lattice system that realizes interaction-induced moir\'e physics in the absence of intrinsic disorder or externally imposed quasiperiodic potentials. The setup consists of two one-dimensional lattices labeled by the spin (or layer) index $\sigma=\uparrow,\downarrow$, with different lattice constants $d_{\sigma}$. A schematic of the lattice configuration is shown in Fig.~\ref{fig: figure1}.

Within the lowest-band approximation, the system is described by a spin-dependent Fermi--Hubbard-type model,
\begin{equation}
\hat{H}
=
-\sum_{\sigma=\uparrow,\downarrow}\sum_{\langle i_\sigma,j_\sigma\rangle}
J_{\sigma}\,\hat{b}^{\dagger}_{i_{\sigma}}\hat{b}_{j_{\sigma}}
+\sum_{i_\uparrow,j_\downarrow}
U_{i_\uparrow,j_\downarrow}\,\hat{n}_{i_\uparrow}\hat{n}_{j_\downarrow}.
\label{eq:Hamiltonian}
\end{equation}
Here $\hat{b}^{\dagger}_{i_{\sigma}}$ ($\hat{b}_{i_{\sigma}}$) creates (annihilates) a fermion on site $i$ in layer $\sigma$ in the Wannier basis, and $\hat{n}_{i_\sigma}\equiv \hat{b}^{\dagger}_{i_{\sigma}}\hat{b}_{i_{\sigma}}$ is the corresponding number operator. The notation $\langle i_\sigma,j_\sigma\rangle$ denotes nearest neighbors within layer $\sigma$. The two layers are coupled solely through the interlayer density--density interaction; the matrix elements $U_{i_\uparrow,j_\downarrow}$ decay rapidly with the interlayer separation, so that introducing a finite spatial cutoff produces negligible quantitative changes.

Equation~\eqref{eq:Hamiltonian} can be implemented in cold-atom platforms. In state-dependent optical lattices, both the hopping amplitudes $J_\sigma$ and the interaction matrix elements can be tuned by varying the lattice wavelengths and depths, while the interspecies coupling can be controlled via Feshbach resonances or by choosing different atomic mixtures. Alternatively, the same Hamiltonian can be realized in optical tweezer arrays, which offer increased flexibility: the lattice constants and relative incommensurability can be engineered directly in real space, and the hopping and interaction strengths can be adjusted largely independently.

\begin{figure}[h]
   \centering
           \includegraphics[width=1\hsize,keepaspectratio]{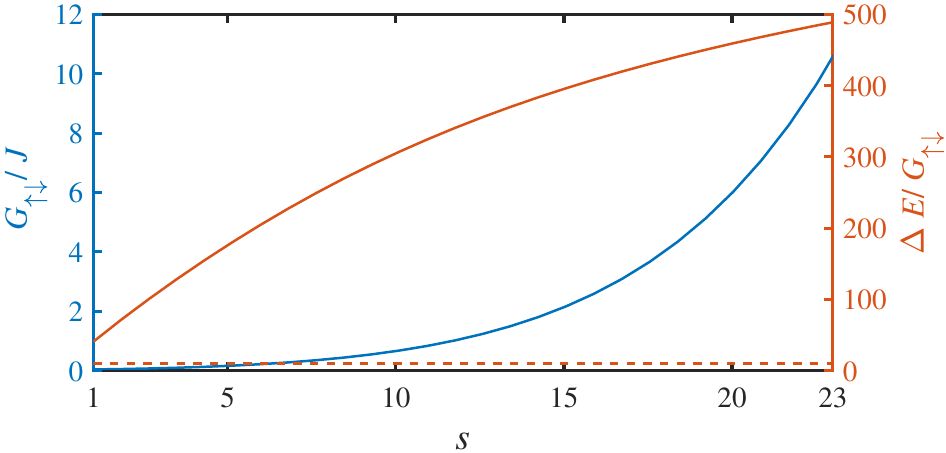}
         \caption{Dependence of the dimensionless interaction strength and band-gap ratio on the lattice depth parameter $s$. The left (blue) axis shows $G_{\uparrow\downarrow}/J$ as a function of $s$, while the right (orange) axis shows the ratio of the single-particle band gap $\Delta E$ (between the first and second bands) to $G_{\uparrow\downarrow}$. The latter quantifies the separation of the lowest band from higher bands and is used to justify the single-band approximation in the parameter regime considered.}
    \label{fig: figure2}
\end{figure}

For concreteness, we briefly summarize the parameter dependence in an optical-lattice implementation. Approximating the Wannier functions by Gaussians, the interlayer interaction takes the form
\begin{equation}
U_{i_\uparrow,j_\downarrow}
=
\frac{g_{\uparrow\downarrow}}{2}\sqrt{\kappa_{\mathrm{eff}}}\,
e^{-\pi\kappa_{\mathrm{ave}} \Delta_{i,j}^2},
\end{equation}
where $\Delta_{i,j}$ is the physical distance between sites $i_\uparrow$ and $j_\downarrow$, $\kappa_{\mathrm{eff}}=\sqrt{\kappa_{\uparrow}\kappa_{\downarrow}}$, and $\kappa_{\mathrm{ave}}=(\kappa_{\uparrow}+\kappa_{\downarrow})/2$. The effective parameters are related to experimentally tunable quantities through
\begin{equation}
g_{\uparrow\downarrow}=\frac{4\pi\hbar^2a_{\uparrow\downarrow}}{\pi\mu\eta^2},\quad
\kappa_{\sigma}=\sqrt{\frac{\mu V_{0}^{\sigma}}{2\hbar^2d_{\sigma}^2}},\quad
\eta=\sqrt{\frac{\hbar}{\mu\omega_{\bot}}},
\end{equation}
where $\omega_{\bot}$ is the transverse confinement frequency, $d_{\sigma}=\lambda_{\sigma}/2$ is the lattice constant set by the optical wavelength $\lambda_{\sigma}$, $a_{\uparrow\downarrow}$ is the interspecies $s$-wave scattering length, $\mu$ is the atomic mass, and $V_{0}^{\sigma}$ is the lattice depth. The two lattices have approximately equal physical length, and the lattice constants of the two layers are related by $d_{\downarrow}/d_{\uparrow} = \beta$. Here $\beta$ is the ratio of Fibonacci numbers, $\beta = F_{n}/F_{n+1}$, where $F_{n+1}$ is defined recursively by $F_{n+1} = F_{n} + F_{n-1}$ with $F_{0} = F_{1} = 1$ \cite{kohmoto1983metal,ingold2002delocalization,wang2016spectral,wang2020realization,prange1984solvable}. For large $n$, $\beta$ asymptotically approaches the Golden Ratio. Throughout this paper, we have chosen $\beta=233/377$. For the lattice numbers $L_{\sigma}$, we first set $L_{\uparrow}$ to be an integer, and then obtain $L_{\downarrow}\approx L_{\uparrow}/\beta$.

In the estimates below we take $a_{\uparrow\downarrow}\approx 5~\mathrm{nm}$ (tunable via magnetic field or by choosing different species; for $^{40}\mathrm{K}$ one has $a_{\uparrow\downarrow}\approx 9~\mathrm{nm}$~\cite{zwergerMottHubbardTransition2003a}), and parameterize the lattice depth as $V_{0}^{\sigma}=sE_{r}^{\sigma}$ with $1\le s\le 23$, where $E_{r}^{\sigma}=h^2/(2\mu\lambda_{\sigma}^2)$ is the recoil energy for layer $\sigma$. In the numerical calculations below, we take $E_r^{\uparrow}$ as the energy unit and also set $\omega_{\bot}=2\pi\times 1~\mathrm{kHz}$. Under the same Gaussian approximation, the nearest-neighbor hopping can be estimated as~\cite{zwergerMottHubbardTransition2003a}
\begin{equation}
J_{\sigma}=1.43\,s^{0.98}e^{-2.07\sqrt{s}}\,E_{r}^{\sigma}.    
\end{equation}
For later convenience, we define the interaction prefactor
\begin{equation}
G_{i_\uparrow,j_\downarrow}\equiv \frac{g_{\uparrow\downarrow}}{2}\sqrt{\kappa_{\mathrm{eff}}},
\end{equation}
and use $G_{\uparrow\downarrow}/J_{\uparrow}$ (denoted as $G_{\uparrow\downarrow}/J$ hereafter for simplicity) as the dimensionless interaction strength. The dependence of $G_{\uparrow\downarrow}/J$ on $s$, together with the corresponding band-gap estimate supporting the single-band approximation, is shown in Fig.~\ref{fig: figure2}.

In the following sections we analyze the localization properties of Eq.~\eqref{eq:Hamiltonian} in two complementary regimes. First, by preparing a regular filling pattern in one layer and quenching its hopping, particles in the other layer experience an interaction-induced quasiperiodic potential with a mosaic-like envelope. Fractional regular fillings further dilute the potential and provide a direct route toward the mosaic Aubry–Andr\'e limit. This can be achieved in optical lattices by deepening one lattice, or more directly in tweezer arrays by increasing the corresponding trap spacing to suppress tunneling. Second, we restore hopping in both layers and investigate how interlayer interactions induce ergodic--critical--MBL behavior as functions of energy and filling in each layer.

\begin{figure}[h]
   \centering
    \includegraphics[width=1\hsize,keepaspectratio]{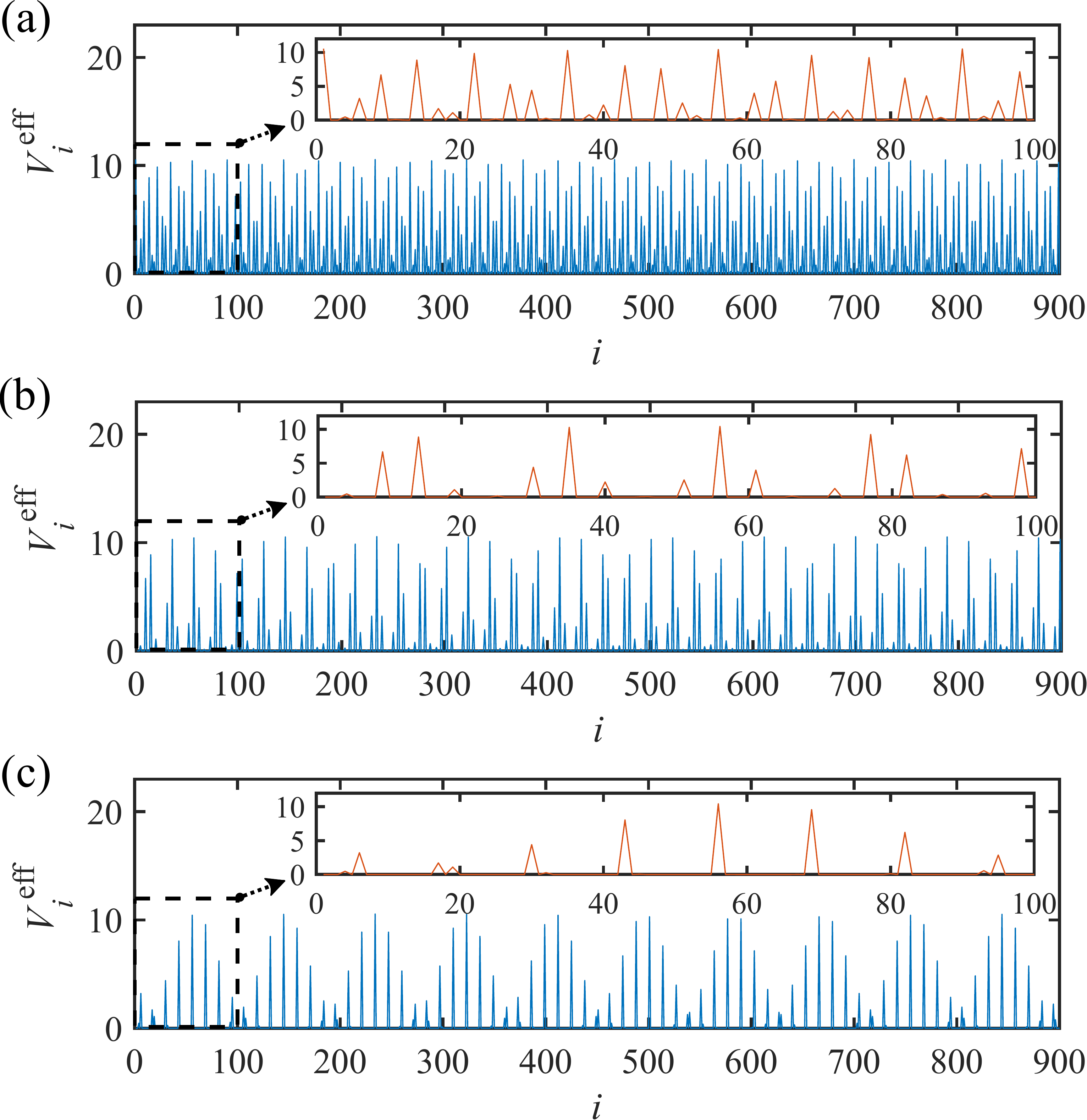}
    \caption{The figure shows the interaction-induced effective potential $V_i^{\mathrm{eff}}$ (in unit of $J_{\uparrow}$) experienced by the upper-layer particle for regular fillings of the lower layer with $s=23$. The panels (a), (b) and (c) correspond to  $\lambda=1$, $\lambda=2$ and $\lambda=3$, respectively. Insets show a magnified view of $V_i^{\mathrm{eff}}$ over lattice sites $i\in[0,100]$, highlighting the mosaic-like envelope of the quasiperiodic potential. Unit filling ($\lambda=1$) yields a dense but modulated envelope, while $\lambda\ge2$ produces a progressively diluted mosaic pattern.}
    \label{fig: figure3}
\end{figure}

\section{INTERACTION-INDUCED MOSAIC POTENTIAL}
For simplicity, we first consider the limit in which the hopping of particles in the lower layer is quenched, i.e., $J_{\downarrow}=0$. In this case, the particle configuration in the lower layer is frozen and acts as a static background. We further focus on the single-particle limit in the upper layer, $N_{\uparrow}=1$. Under these conditions, the particle in the upper layer experiences an effective external potential generated entirely by its interaction with particles in the lower layer.

Specifically, the effective potential at site $i$ of the upper layer is given by
\begin{equation}
V_i^{\mathrm{eff}}=\sum_j U_{i_\uparrow,j_\downarrow}\,
\langle \hat{n}_{j_\downarrow}\rangle .
\end{equation}
Because the two layers have different lattice constants, this interaction-induced potential is generally quasiperiodic when the lower layer is populated. Moreover, its detailed spatial structure depends sensitively on the particle distribution in the lower layer, encoded in $\langle \hat{n}_{j_\downarrow}\rangle$.

To characterize the nature of the eigenstates, we employ the fractal dimension of the wave function,
\begin{equation}
\Gamma(m)
=
-\lim_{L\rightarrow\infty}
\left[
{\ln\!\left(\sum_{j=1}^{L} |\Psi_{m,j}|^4\right)}/{\ln L}
\right],
\end{equation}
where $\Psi_{m,j}$ denotes the amplitude of the $m$th eigenstate on site $j$. In practice, we take $L=L_{\uparrow}=900$ and $L_{\downarrow}=1456$ for numerical calculations. Extended states are characterized by $\Gamma\to 1$, while localized states satisfy $\Gamma\to 0$ \cite{wangOneDimensionalQuasiperiodicMosaic2020,evers2008anderson}.

\begin{figure}[h]
   \centering
    \includegraphics[width=1\hsize]{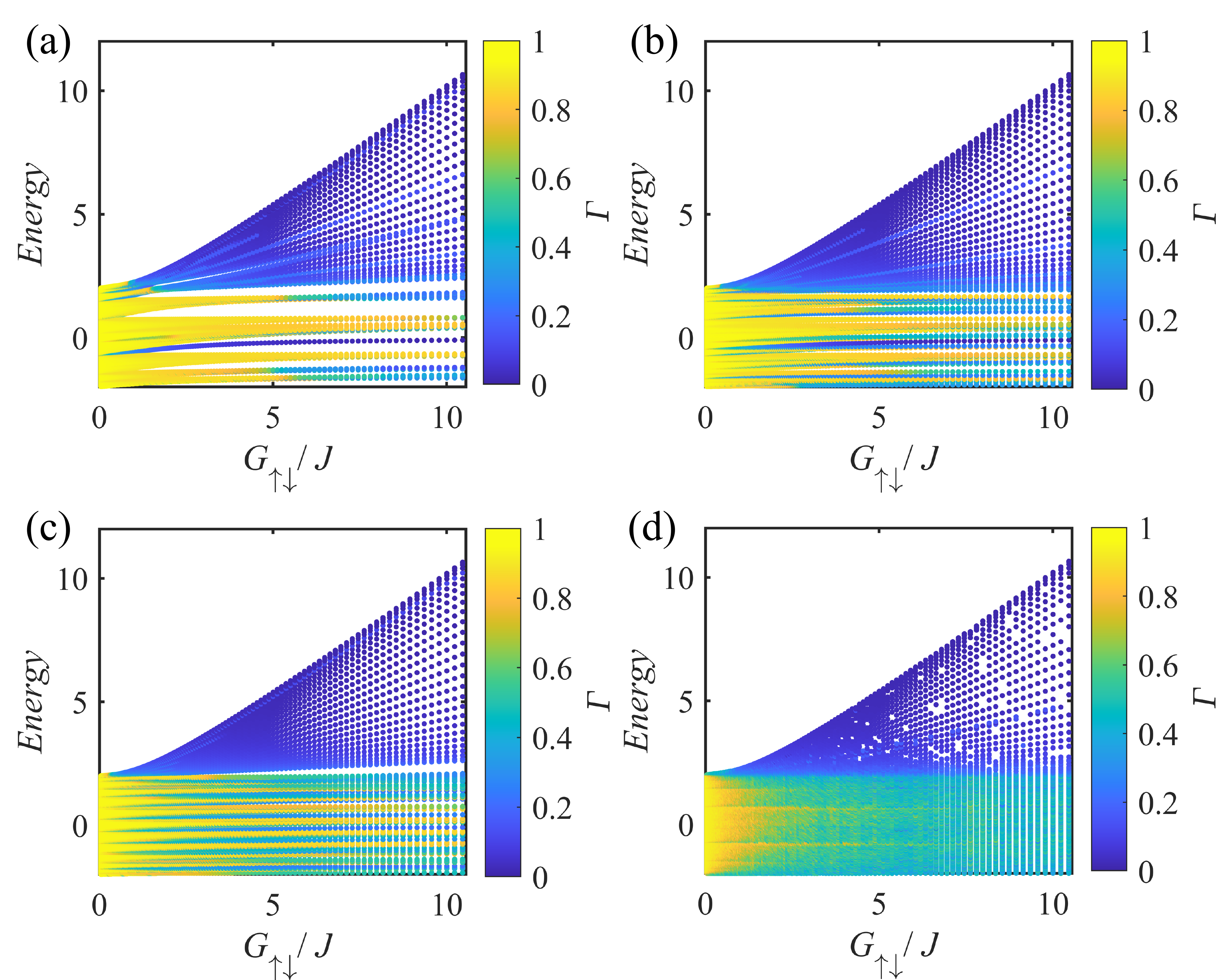}
       \caption{Phase diagrams in the quenched-layer (mosaic) limit with $N_{\uparrow}=1$ and $J_{\downarrow}=0$. The vertical axis shows the single-particle eigenenergy $E$ of the upper layer in units of $J_{\uparrow}$. In all panels, $L_{\uparrow}=900$. Panels (a), (b) and (c) correspond to regular (mosaic) fillings in the lower layer with $N_{\downarrow}/L_{\downarrow}=1$ ($\lambda=1$), $N_{\downarrow}/L_{\downarrow}=1/2$ ($\lambda=2$) and $N_{\downarrow}/L_{\downarrow}=1/3$ ($\lambda=3$), respectively. Panel (d) shows the corresponding phase diagram for an average filling of $N_{\downarrow}/L_{\downarrow}=1/3$ with random configuration in the lower layer. The color indicates the fractal dimension $\Gamma$ of the eigenstates (blue: localized, $\Gamma\!\approx\!0$; yellow: extended, $\Gamma\!\approx\!1$), revealing multiple mobility edges for the regular fillings and their disappearance for random fillings.}
    \label{fig: figure4}
\end{figure}

We first consider a regular filling pattern in the lower layer,
\begin{equation}
\langle \hat{n}_{j_\downarrow} \rangle =
\begin{cases}
1, & j=p\lambda, \\
0, & \text{otherwise},
\end{cases}
\end{equation}
where $\lambda$ is a positive integer and $p=1,2,\ldots$. This corresponds to an evenly spaced filling in which every $\lambda$th site is occupied while all other sites are empty. We focus on three representative cases, $\lambda=1$, $\lambda=2$ and $\lambda=3$. For these cases, the effective potential experienced by the upper-layer particle is shown in Fig.~\ref{fig: figure3}. The envelope of the potential clearly exhibits a mosaic-like structure, characteristic of quasiperiodic mosaic models.

Such regular filling configurations can be realized experimentally in several ways. In optical lattices, one may strongly increase the lattice depth of the lower layer so that its tunneling timescale becomes much longer than that of the upper layer. Alternatively, optical tweezer arrays with Rydberg atoms provide a natural route to enforcing regular filling patterns via the Rydberg blockade mechanism. For example, if the blockade radius satisfies $d_{\downarrow}<r_b<2d_{\downarrow}$, one obtains an evenly distributed filling $N_{\downarrow}/L_{\downarrow}=1/2$ corresponding to $\lambda=2$, while a filling $N_{\downarrow}/L_{\downarrow}=1/3$ with $\lambda=3$ can be realized when $2d_{\downarrow}<r_b<3d_{\downarrow}$. 

Under these conditions, the problem reduces to a noninteracting single-particle model for the upper layer. We compute the eigenstates and eigenenergies by direct diagonalization and evaluate the fractal dimension $\Gamma$ as a function of energy and the interlayer interaction strength $G_{\uparrow\downarrow}/J$. The resulting phase diagrams for $\lambda=1$, $\lambda=2$ and $\lambda=3$ are shown in Figs.~\ref{fig: figure4}(a), \ref{fig: figure4}(b) and \ref{fig: figure4}(c), respectively. All three regular fillings, including unit filling $\lambda=1$, exhibit mosaic-model behavior, most notably the presence of multiple mobility edges that allow states within certain energy ranges to remain extended even at strong coupling. Fractional regular fillings ($\lambda\ge2$) offer the clearest connection to the mosaic–Aubry–Andr\'e picture of diluted active sites, whereas $\lambda=1$ produces a dense yet still quasiperiodic interaction-induced potential whose non-sinusoidal structure can generate mobility edges. The number and location of mobility edges depend on $\lambda$, while the detailed structure differs from standard mosaic models due to the specific interaction-induced form of $V_i^{\mathrm{eff}}$.

Finally, to demonstrate that the appearance of multiple mobility edges relies on the regularity of the filling pattern, we also consider a random distribution of particles in the lower layer. The corresponding phase diagram, shown in Fig.~\ref{fig: figure4}(d), no longer exhibits multiple mobility edges, indicating that this feature is unique to the interaction-induced mosaic potential generated by regular fillings.

\section{Interaction induced MBL}

Next, we restore hopping in both layers and study the interaction-induced MBL properties of the full model. Before presenting the numerical results, we summarize the diagnostics used to distinguish the ergodic (thermal), critical, and MBL regimes. Throughout this work, phases are identified primarily through participation entropies in Hilbert space, and are further corroborated by spectral (level) statistics.

(i) Participation entropies (PE). We characterize localization in the many-body Hilbert space using the participation entropies $S_q^P$. For an eigenstate $|n\rangle$ expanded in the computational basis with probabilities $p_i$, the PEs are defined as \cite{luitzManybodyLocalizationEdge2015,bell1972dynamics,wegner1980inverse,rodriguez2011multifractal,luitz2014universal,luitz2014participation}
\begin{equation}
S^P_q(|n\rangle)=
\begin{cases}
-\sum_i p_i\ln(p_i), & q=1,\\[4pt]
\frac{1}{1-q}\ln\!\left(\sum_i p_i^q\right), & q\neq 1.
\end{cases}
\label{eq:PE}
\end{equation}
Let $\dim\mathcal{H}$ denote the Hilbert-space dimension. In the ergodic regime, eigenstates are extended over a finite fraction of the basis states, leading to the scaling
\begin{equation}
S^P_q \simeq a_q\ln(\dim\mathcal{H}),\qquad a_q\simeq 1 \ \ (\forall q).
\end{equation}
In contrast, in the localized regime the support of eigenstates in the computational basis is strongly reduced. A commonly used form is \cite{luitzManybodyLocalizationEdge2015}
\begin{equation}
S^P_q \simeq a_q\ln(\dim\mathcal{H}) + l_q\ln\!\big[\ln(\dim\mathcal{H})\big],
\end{equation}
which grows with system size much more slowly than in the ergodic phase, with an effective coefficient $a_q\ll 1$ (or, asymptotically, $a_q=0$ with $l_q>0$). In the following, we use the coefficient $a_1$ extracted from $S_1^P$ as an operational indicator of ergodic versus localized behavior in the system sizes accessible to exact diagonalization.

(ii) Level statistics. As an independent probe, we examine spectral statistics within the framework of random-matrix theory. In the ergodic regime, the level spacing distribution exhibits Gaussian orthogonal ensemble (GOE)-type level repulsion, while the localized regime approaches Poisson statistics due to uncorrelated levels. A widely used and numerically stable diagnostic is the ratio of consecutive level spacings \cite{oganesyanLocalizationInteractingFermions2007,wangManybodyGroundState2016,luitzManybodyLocalizationEdge2015,bauerAreaLawsManybody2013,palManybodyLocalizationPhase2010,cuevas2012level,laumann2014many}
\begin{equation}
r^{(n)}=\frac{\min(\delta^{(n)},\delta^{(n+1)})}{\max(\delta^{(n)},\delta^{(n+1)})},
\qquad
\delta^{(n)}=E_n-E_{n-1},
\label{eq:r}
\end{equation}
whose disorder/sample average interpolates between $r_{\mathrm{GOE}}=0.5307$ and $r_{\mathrm{Poisson}}=2\ln 2-1\simeq 0.3863$ \cite{atas2013distribution,luitz2015many}.

We set the lattice sizes to $L_{\uparrow}=16$ and $L_{\downarrow}=26$, whose ratio is chosen from consecutive Fibonacci numbers to approximate the golden ratio and thus maximize quasiperiodicity at finite size. We perform exact diagonalization with open boundary conditions. For each eigenenergy $E$, we define the normalized energy
$\epsilon=\left(E-E_0\right)/\left(E_{\mathrm{max}}-E_0\right)$,
where $E_0$ and $E_{\mathrm{max}}$ are the ground-state and highest excited-state energies, respectively.

\begin{figure}[h]
   \centering
    \includegraphics[width=1\hsize,keepaspectratio]{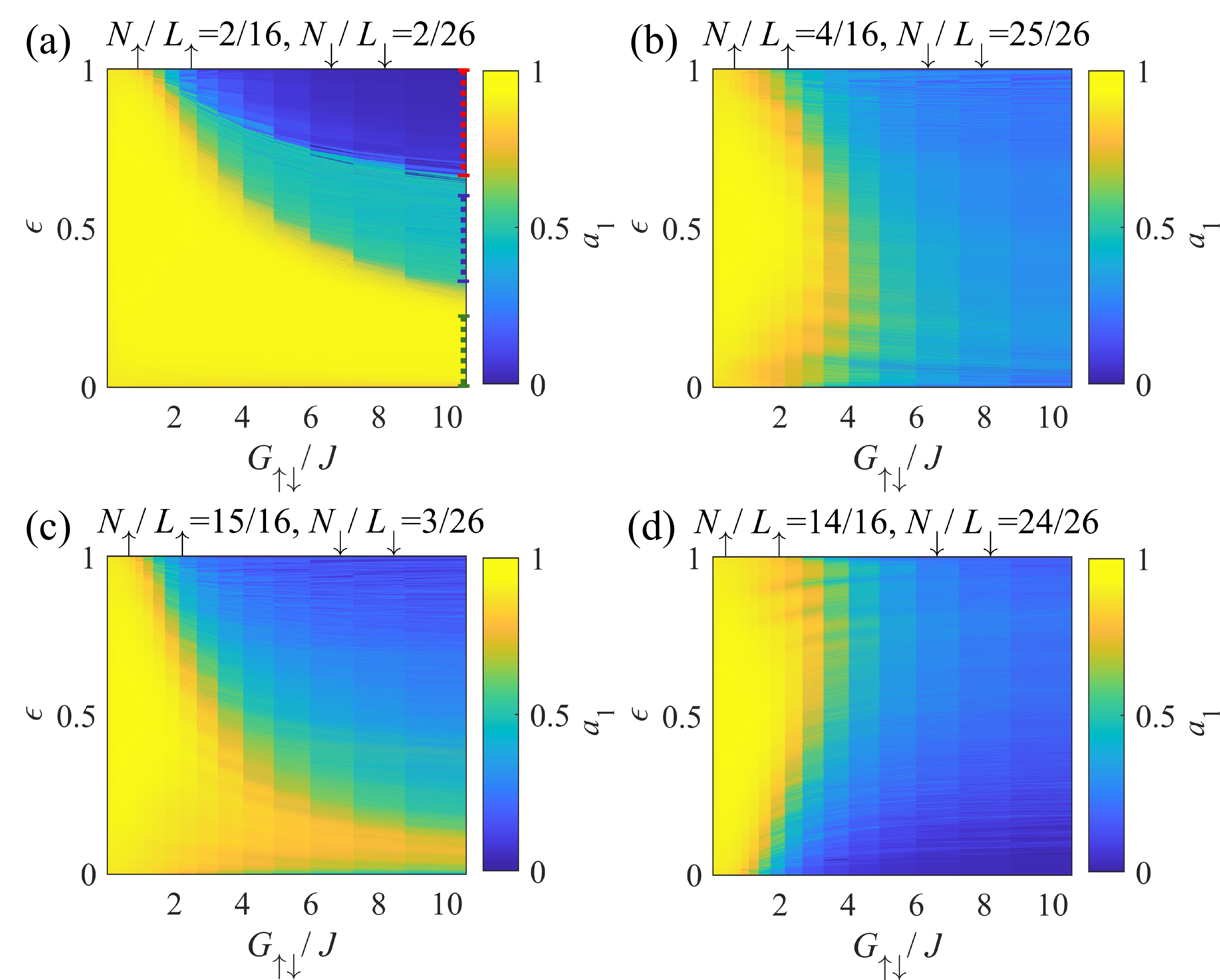}
       \caption{Phase diagrams under different filling conditions. Panels (a)--(d) correspond to
       $N_{\uparrow}/L_{\uparrow}=2/16,\,4/16,\,15/16,\\\,14/16$ and $N_{\downarrow}/L_{\downarrow}=2/26,\,25/26,\,3/26,\,24/26$, respectively.
       Yellow indicates the ergodic (thermal) regime, light blue denotes an intermediate (critical) regime, and dark blue indicates the MBL regime. The colorbar shows the participation coefficient $a_1$.}
    \label{fig: figure5}
\end{figure}

Using the eigenstates obtained from exact diagonalization, we compute the participation entropy for each energy level and analyze its dependence on the interlayer interaction strength $G_{\uparrow\downarrow}/J$, thereby obtaining the phase diagrams shown in Fig.~\ref{fig: figure5}.
As illustrated in Fig.~\ref{fig: figure5}, contrary to the conventional expectation that interactions tend to delocalize, here localization is induced by the interlayer interaction. Increasing $G_{\uparrow\downarrow}/J$ can drive a transition from an ergodic spectrum to an MBL spectrum. This behavior is qualitatively distinct from standard quasiperiodic MBL models, where the quasiperiodic potential exists already at the single-particle level: in our system, the two layers decouple into two periodic lattices when $G_{\uparrow\downarrow}=0$, and quasiperiodicity emerges only through interlayer interactions.

The phase diagrams exhibit a pronounced dependence on filling. At low filling in one or both layers, a substantial portion of the spectrum remains ergodic even at strong interlayer interactions. As the filling increases, the boundary between the ergodic and MBL regimes evolves into a dome-like structure, as seen in Fig.~\ref{fig: figure5}(d). This trend can be understood as follows: at higher fillings, particles more frequently encounter interlayer interaction, and the effective potential experienced by particles becomes closer to a fully developed quasiperiodic landscape with a reduced density of ``mosaic'' (effectively interaction-free) configurations. In contrast, at low fillings, interactions are absent on a large fraction of sites/configurations, leading to a high density of mosaic regions and hence a stronger tendency toward delocalization, as in Fig.~\ref{fig: figure5}(a).

A particularly intriguing case is shown in Fig.~\ref{fig: figure5}(a), where both layers contain only two particles. The spectrum exhibits pronounced mobility edges separating ergodic, critical, and MBL regimes: as the energy increases, eigenstates evolve from ergodic to critical and then to localized. Physically, higher-energy states typically involve configurations with larger interlayer interaction energy, which requires particles in opposite layers to approach each other more closely; only in such configurations does the interaction-induced quasiperiodic potential become effective, thereby favoring localization at higher energies.

\begin{figure}[h]
   \centering
           \includegraphics[width=1\hsize,keepaspectratio]{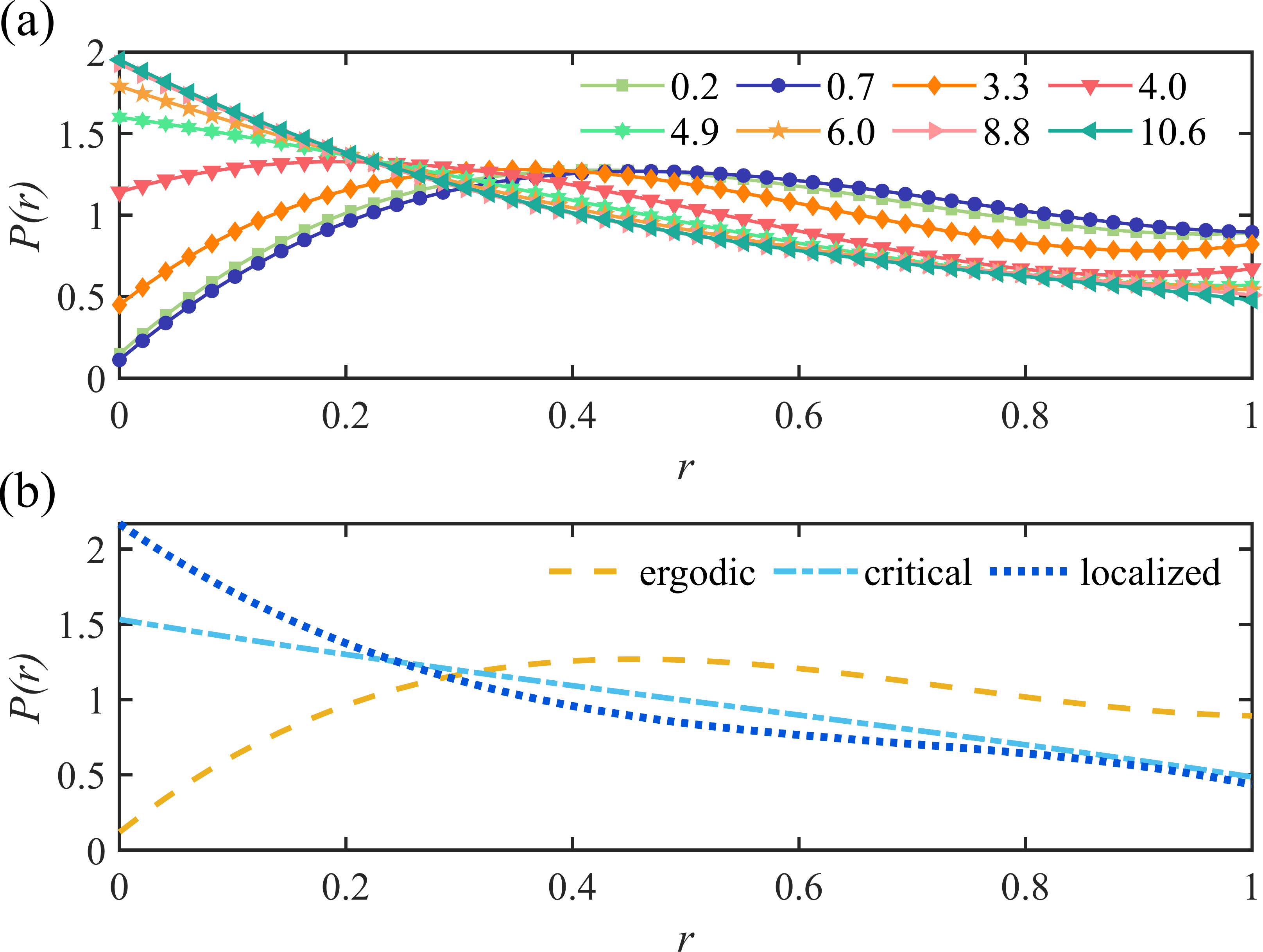}
       \caption{Characterization of localization using level statistics. Panel (a) corresponds to Fig.~\ref{fig: figure5}(d). Curves of different colors correspond to increasing interlayer interaction strengths $G_{\uparrow\downarrow}/J$ (indexed by $s=6,10,17,18,19,20,22,23$). In panel (b), the yellow dashed line, light blue dash-dotted line, and dark blue dotted line show $P(r)$ computed within the energy windows $\epsilon\in(0,0.2)$, $(0.3,0.6)$, and $(0.64,1)$, respectively, for $s=23$. These windows correspond to the green, blue, and red dashed intervals marked in Fig.~\ref{fig: figure5}(a), respectively.}
    \label{fig: figure6}
\end{figure}

To validate the phase identification based on participation entropy, we further compute level statistics for representative parameters.
We choose several values of $G_{\uparrow\downarrow}/J$ (indexed by different $s$) corresponding to vertical cuts in Fig.~\ref{fig: figure5}(d) and Fig.~\ref{fig: figure5}(a), and show the resulting level spacing distributions $P(r)$ in the upper and lower panels of Fig.~\ref{fig: figure6}, respectively. Clearly, $P(r)$ shows features that distinguish phases with different localization properties. For the high-filling case in Fig.~\ref{fig: figure5}(d), there is a transition between ergodic and MBL phases with increase of $G_{\uparrow\downarrow}/J$, with the transition occurring near $G_{\uparrow\downarrow}/J\sim 4.0$: when $G_{\uparrow\downarrow}/J<1$, the statistics are GOE-like and exhibit level repulsion ($P(r)\to 0$ as $r\to 0$), while for large $G_{\uparrow\downarrow}/J$ the distribution approaches the Poisson form, with good agreement at $G_{\uparrow\downarrow}/J=8.8$ and $10.6$. This confirms the PE-based identification of ergodic and localized regimes.

In contrast, the low-filling case corresponding to Fig.~\ref{fig: figure5}(a) displays strong energy dependence. Although some high-energy eigenstates exhibit clear localization signatures at strong interactions, low-energy states remain ergodic. Moreover, the PE shows plateau-like behavior across different energy windows at fixed $G_{\uparrow\downarrow}/J$, suggesting distinct regimes within the spectrum. To resolve this, we select three energy intervals indicated by the dashed lines in Fig.~\ref{fig: figure5}(a) and compute their level statistics, as shown in Fig.~\ref{fig: figure6}(b). The high-energy window is Poisson-like (localized), the low-energy window remains GOE-like (ergodic), and the intermediate window exhibits critical behavior. Taken together, these results demonstrate the coexistence of ergodic, critical, and many-body localized regimes in the same model, separated by interaction-induced mobility edges \cite{PhysRevLett.126.080602,zhou2025fundamental,li2026multifractal}.

In principle, a complete characterization of the phase boundaries requires finite-size scaling. Here this is particularly challenging because the phase diagram depends sensitively on the filling factors in both layers; maintaining fixed fillings while increasing system size imposes additional constraints on accessible sizes. Combined with the exponential growth of the Hilbert space in exact diagonalization, this limits the range of sizes available for scaling analysis. We therefore focus on the largest system sizes within our computational capability and use the consistency between participation entropies and level statistics to support our conclusions.

\section{Mapping to higher-dimensional non-interacting model}

To fully understand the results presented in Fig. \ref{fig: figure5},
we present an exact mapping that transforms the interacting many-body Hamiltonian in Eq.~(1) into a single-particle model on a structured higher-dimensional lattice. This mapping holds for any fixed particle numbers $N_{\sigma}$ and lattice sizes $L_{\sigma}$.

Let us denote the positions of $N_\sigma$ fermions in layer $\sigma$ by the tuple $(x_1, x_2, \dots, x_{N_\sigma})$. Since these particles are identical fermions, this property can be imposed through the constrain in the lattice positions, $1 \leq x_1 < x_2 < \cdots < x_{N_\sigma} \leq L_\sigma$. Geometrically, this set forms a $N_\sigma$-dimensional simplex region within a $N_\sigma$-dimensional hypercubic lattice, bounded by the constraints $x_i<x_{i+1}$. We denote this configuration space as $\mathcal{C}_\sigma$, which contains $C_{L_{\sigma}}^{N_{\sigma}}=L_{\sigma}!/[N_{\sigma}!(L_{\sigma}-N_{\sigma})!]$ lattice sites.

From this, we construct a graph $G_\sigma$ whose vertices are the configurations in $\mathcal{C}_\sigma$. Two vertices are connected by an edge if their corresponding configurations differ by moving exactly one fermion to a neighboring site, with hopping amplitude $-J_\sigma$. This yields a sparse, structured graph that inherits the geometry of the simplex region.

For the two-layer system, the full Hilbert space is the tensor product $\mathcal{C}_\uparrow \otimes \mathcal{C}_\downarrow$. The corresponding graph is the Cartesian product $G = G_\uparrow \times G_\downarrow$, whose vertices are pairs of configurations $(\alpha,\beta)$ and whose edges correspond to single-fermion hops within either lattice.

\begin{figure}[h]
   \centering
           \includegraphics[width=1\hsize,keepaspectratio]{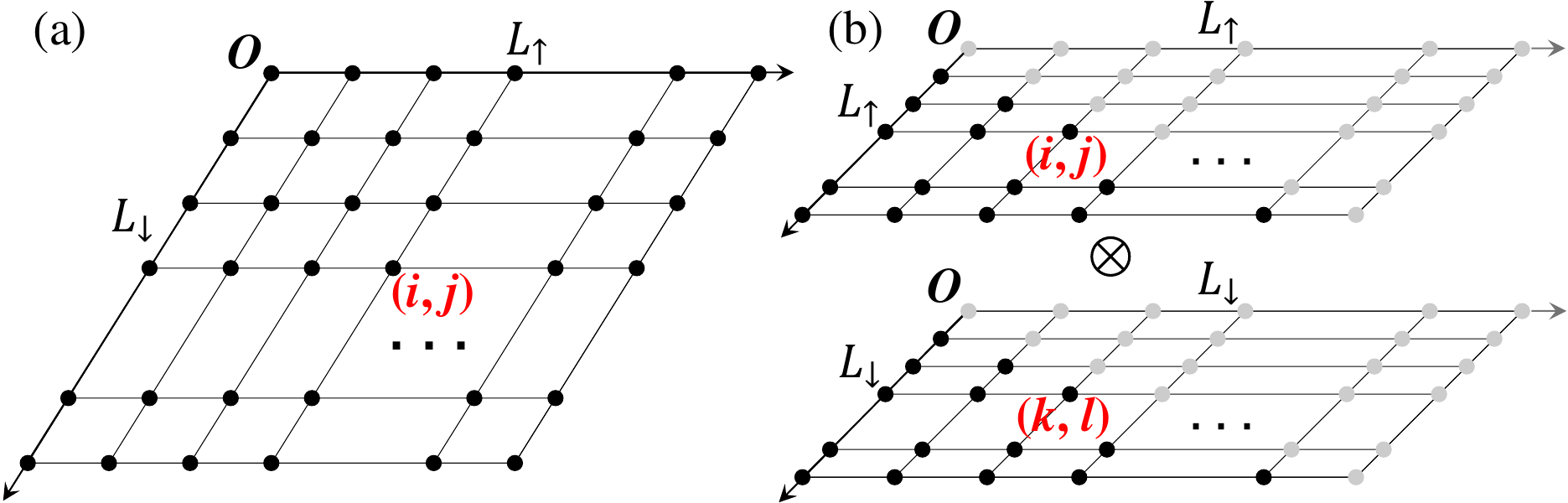}
       \caption{Structure of the mapped non-interacting lattice. (a) $N_{\uparrow}=N_{\downarrow}=1$: the graph $G$ reduces to a rectangular lattice of size $L_{\uparrow}\times L_{\downarrow}$. (b) $N_{\uparrow}=N_{\downarrow}=2$: each layer's configuration space is a 2-simplex (triangular region with black dots) embedded in an $L_\sigma\times L_\sigma$ square lattice, and the full graph is their Cartesian product.}
    \label{fig: mapping}
\end{figure}

In this basis, the many-body Hamiltonian becomes exactly a single-particle tight-binding model on $G$:
\begin{equation}
\begin{aligned}
\hat{H} = &\sum_{(\alpha,\beta)} V_{\alpha\beta} |\alpha,\beta\rangle\langle\alpha,\beta| + \sum_{\text{edges}} \Bigl( t_{(\alpha,\beta)\to(\alpha',\beta)} |\alpha',\beta\rangle\langle\alpha,\beta| \\
& + t_{(\alpha,\beta)\to(\alpha,\beta')} |\alpha,\beta'\rangle\langle\alpha,\beta| + \text{h.c.} \Bigr),
\end{aligned}
\label{eq:graph_ham}
\end{equation}
where $V_{\alpha\beta}$ just comes from interlayer interactions
$V_{\alpha\beta} =  \sum_{i\in \alpha, j\in \beta} U_{i\uparrow j\downarrow}$,
and hopping amplitudes $t$ are $-J_\uparrow$ or $-J_\downarrow$. Equation~\eqref{eq:graph_ham} describes a single particle hopping on the graph $G$ with onsite potentials $V_{\alpha\beta}$. Thus, the interacting many-body problem maps exactly to a non-interacting single-particle model on a higher-dimensional graph.

\begin{figure}[h]
   \centering
           \includegraphics[width=1\hsize,keepaspectratio]{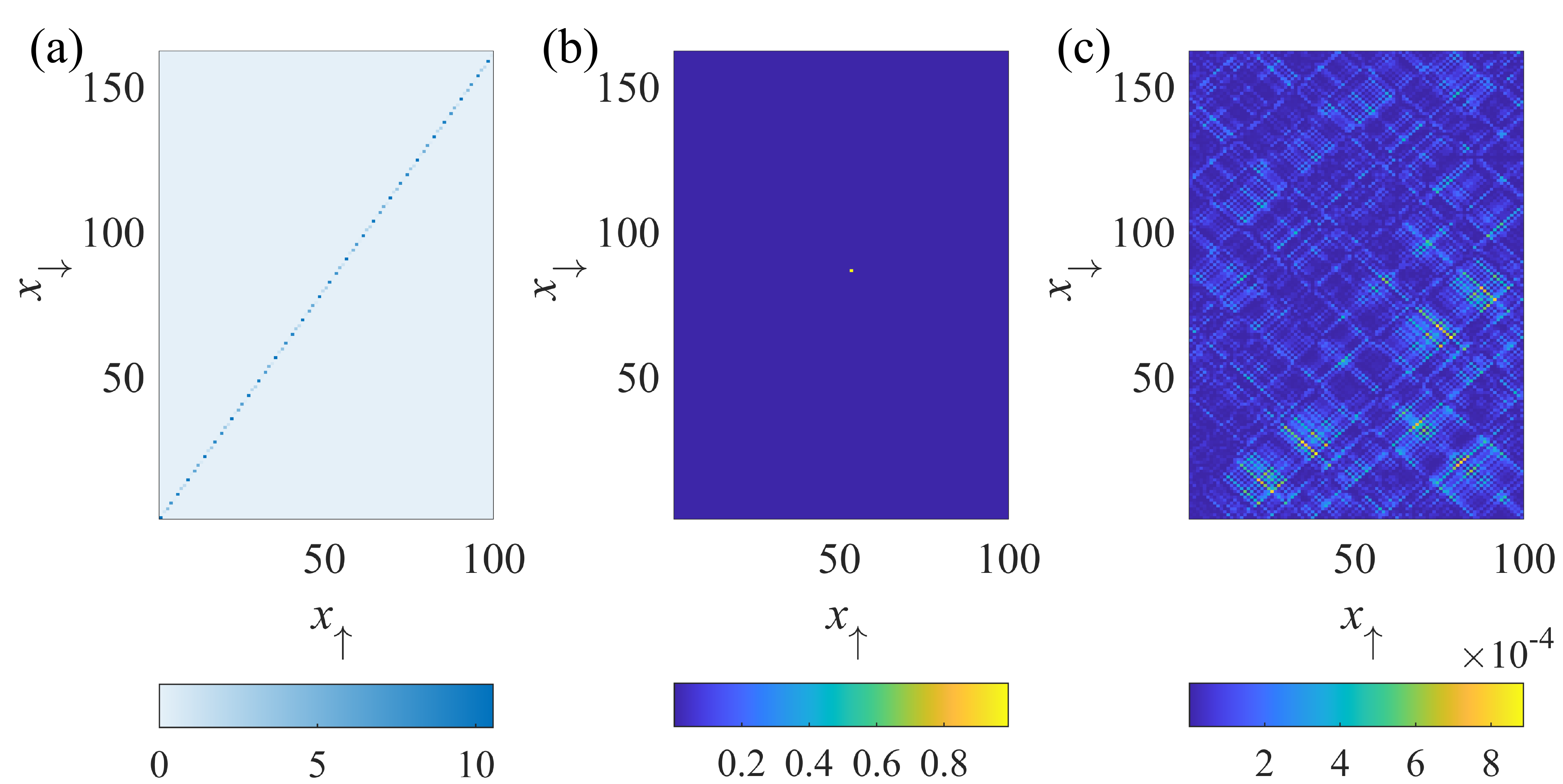}
       \caption{Mapped onsite potential and representative eigenstates for $N_{\uparrow}=N_{\downarrow}=1$. (a) Onsite potential $V_{i,j}$ on the $L_\uparrow\times L_\downarrow$ lattice and the wavefunction localizes precisely to a single point on the diagonal, where $V_{i,j}=U_{i\uparrow j\downarrow}$. (b) Probability distribution of a high-energy eigenstate. (c) Probability distribution of the ground state. Parameters: $s=23$, $L_\uparrow=100$ and $L_\downarrow=160$.}
    \label{fig: figure9}
\end{figure}

To illustrate this mapping, we examine two specific cases. For one particle in each layer ($N_\uparrow=N_\downarrow=1$), $\mathcal{C}_\sigma=\{1,\dots,L_\sigma\}$ and $G_\sigma$ is a one-dimensional lattice. The product graph $G$ is a two-dimensional lattice of size $L_\uparrow\times L_\downarrow$ [Fig.~\ref{fig: mapping}(a)], where the onsite potential at site $(i,j)$ is simply $V_{i,j}=U_{i\uparrow j\downarrow}$ [Fig.~\ref{fig: figure9}(a)]. Because $U_{i\uparrow j\downarrow}$ is short ranged, $V_{i,j}$ is appreciable only when the two particles are spatially close, so the nonzero potential forms a narrow strip near the diagonal. The resulting model is therefore a two-dimensional mosaic potential problem. For the typical onsite potential shown in Fig.~\ref{fig: figure9}(a), the probability distributions of the ground state and a high-energy eigenstate within the $L_\uparrow\times L_\downarrow$ space are displayed in Fig.~\ref{fig: figure9}(b) and Fig.~\ref{fig: figure9}(c), respectively. The ground-state wave function spreads across the entire space. As the energy increases, the wave function becomes progressively more localized, eventually concentrating at a single point along the diagonal.

For two particles in each layer ($N_\uparrow=N_\downarrow=2$), each configuration is an ordered pair $(i,j)$ with $i<j$. Geometrically, $\mathcal{C}_\sigma$ is the triangular region of an $L_\sigma\times L_\sigma$ square lattice excluding the diagonal $i=j$. The full graph $G=G_\uparrow\times G_\downarrow$ is then the Cartesian product of two such triangles [Fig.~\ref{fig: mapping}(b)], i.e., a structured four-dimensional lattice. The onsite potentials $V_{\alpha\beta}$ again encode the interlayer interaction energy between the two configurations.

This mapping directly explains the main features of Fig.~\ref{fig: figure5}. First, the effective potentials $V_{\alpha\beta}$ inherit quasiperiodic structure from the incommensurate lattice constants: although each layer is periodic in isolation, the relative positions of particles across the two lattices generate configuration-dependent interaction energies. Increasing the interlayer interaction strength amplifies the fluctuations of $V_{\alpha\beta}$, thereby enhancing the effective disorder in the mapped model and driving a transition from extended to localized states on the graph---which corresponds to MBL in the original system.

Second, for small particle numbers (or low fillings), the nonzero values of $V_{\alpha\beta}$ occupy only a sparse subset of configuration space because interactions are short ranged. Large regions of the graph are effectively ``mosaic'' in configuration space (nearly zero potential), naturally favoring extended or critical behavior and explaining the prominent intermediate regimes.

Finally, both mobility edges and the filling dependence become transparent in the graph picture. Different energy sectors probe different regions of configuration space: higher-energy eigenstates typically involve more frequent interlayer contacts and thus experience stronger effective potential variations, making them more prone to localization. Likewise, at higher fillings a larger fraction of configurations carry nonzero interaction energy, producing a more homogeneous quasiperiodic landscape on $G$ and stabilizing localization, whereas at low fillings the abundance of mosaic configurations sustains delocalization or criticality over wide energy ranges.

\section{discussion and conclusion}

The interaction-induced localization mechanism identified here is not restricted to one-dimensional systems and can be naturally generalized to twisted two-dimensional lattices. In such geometries, a relative twist angle generates long-wavelength moir\'e patterns in the interlayer interaction energy, which act as an emergent quasiperiodic potential. We therefore expect interaction-induced single-particle and many-body localization to persist in two dimensions, with richer phase structures due to enhanced connectivity. These predictions can be directly tested in twisted bilayer optical lattice experiments, where lattice geometry, twist angle, and interlayer coupling are highly tunable. Localization may be probed through quench dynamics, expansion measurements, and spectral diagnostics, with particle filling providing a clear experimental control parameter.

Although our analysis focused on fermions, the same interaction-induced moir\'e mechanism should apply to bosonic systems. In that case, the emergent quasiperiodic landscape generated by interlayer interactions is expected to favor a disorder-free Bose glass regime, characterized by suppressed global phase coherence and finite compressibility, without requiring any extrinsic randomness. Moreover, beyond optical lattices, the Hamiltonian considered here can be implemented with even greater flexibility in optical tweezer arrays, where incommensurate geometries can be engineered directly in real space and where hopping and interaction strengths can be tuned largely independently. This enables controlled access to both the mosaic limit (by freezing one layer) and the fully dynamical regime (with hopping in both layers) within the same experimental architecture.

In summary, we have demonstrated that interparticle interactions alone can induce localization in moir\'e lattice systems, even in the complete absence of intrinsic disorder or externally imposed quasiperiodic potentials. Using a spin-dependent bilayer lattice with incommensurate lattice constants, we showed that interactions can generate effective mosaic potentials, multiple mobility edges, and interaction-driven transitions between ergodic, critical, and many-body localized phases. An exact mapping to a noninteracting model on a higher-dimensional structured graph provides a unified physical interpretation of these phenomena. Our results establish interaction-induced moir\'e lattices as a versatile and experimentally accessible platform for studying localization physics and open new directions for exploring disorder-free MBL in both fermionic and bosonic quantum systems.

\begin{acknowledgements}
We acknowledge valuable suggestions given by Zhi Li and Chang Li. Y. Y. and Q. Z. are supported by the National Key Research and Development Program of China (Grant No. 2022YFA1405304), the National Natural Science Foundation of China (Grant No. 12574193) and the Guangdong Provincial Quantum Science Strategic Initiative (Grant Nos. GDZX2401002, GDZX2501003). L. Y. is supported by the Natural Science Foundation of China (No. 12375021), the Zhejiang Provincial Natural Science Foundation of China (No. LD25A050002), and the National Key Research and Development Program of China (No. 2022YFA1404203). Z. X. is supported by the National Natural Science Foundation of China (Grants No. 12375016 and No. 12461160324), and the Beijing National Laboratory for Condensed Matter Physics (Grant No. 2023BNLCMPKF001).
\end{acknowledgements}

DATA AVAILABILITY.---
The data that support the findings of this article are openly available.

\appendix


\bibliography{many_body_localization}
\end{document}